\documentclass[12pt]{article}
\usepackage{graphicx}
\usepackage{hyperref}
\usepackage{dcolumn}
\usepackage[super,sort&compress]{natbib}
\usepackage{amssymb}

\newcolumntype{d}[1]{D{.}{.}{#1}}

\title{Physical exact conditions as regularizers for exchange-correlation in solids and surface chemistry}

\author{Johannes Voss\\
SUNCAT Center for Interface Science and Catalysis\\ SLAC National Accelerator Laboratory, Menlo Park, CA, USA\\
\href{mailto:vossj@slac.stanford.edu}{vossj@slac.stanford.edu}}

\begin{document}
\maketitle


\begin{abstract}
Density functional theory (DFT) often is the method of choice for simulating the electronic properties of extended solids and surfaces from first principles due to a favorable compromise between accuracy and computational cost. In the field of heterogeneous catalysis, DFT is indispensable for deriving mechanistic insights and understanding trends in surface reactivity. The accuracy of DFT for surface reaction energetics depends strongly on the exchange-correlation (XC) approximation. We show here that optimization of such XC functionals for surface binding energies can lead to a worse description of surface reaction barriers, unless important physical exact conditions are fulfilled.
\end{abstract}

\section{Introduction}

Electronic structure methods are essential for predictions of materials thermodynamics, electronic properties, and chemical reaction dynamics. For molecular systems, very high accuracies in predictions can be achieved with quantum chemistry methods. For solids and surfaces, the use of such wave-function based methods is problematic, and density functional theory (DFT) is the standard approach for atomistic simulations.

For such simulation of solids, it has been shown that incorporation of exact limits from the homogeneous electron gas model system into the exchange-correlation (XC) density functional leads to improved predictions. At the generalized-gradient approximation (GGA) level of XC functionals, tuning the low-reduced gradient behavior of the PBE\cite{PBE} functional to restoring the second order gradient expansion of the free electron gas exchange energy and correspondingly adjusted correlation improves lattice constants and surface energies, resulting in the PBEsol\cite{PBESOL} functional. These improvements in bulk elastic predictions come at the price of a worse description of bulk cohesive and other reaction energies. Meta-GGAs, that in addition to reduced density gradient information also are parametrized against Kohn-Sham kinetic energy densities can partially mitigate this trade-off.\cite{TPSS} In particular, the SCAN meta-GGA, that fulfills several physical constraints and norms yields both accurate bulk elastic and cohesive properties.\cite{SCAN}

In heterogeneous catalysis, predicting and understanding how the catalyst surfaces lower reaction barriers and how barriers for competing reactions---desired and undesired---compete is key to developing more performant catalysts. As the extended catalyst surfaces typically require that atomistic simulations resort to DFT, accuracies in surface binding energies and reaction barriers are often limited. Advances in XC functional development targeting surface binding energies of relevance to heterogeneous catalysis are often empirical in nature,\cite{BEEFVDW} while the above mentioned, strongly constrained approaches with improved performance for bulk properties typically predict too strong binding to transition-metal surfaces and perform worse than simpler GGA functionals.\cite{ADS41}

Here, we show that empirical approaches that fit against experimental benchmarks for surface binding and bulk properties but also are constrained to fulfill physical XC constraints yield good surface barriers on average, while empirical approaches without such constraints show a worse description of such barriers than simple GGAs.

\section{Approach}

\begin{figure}
 \centering
        \includegraphics[width=0.2\textwidth]{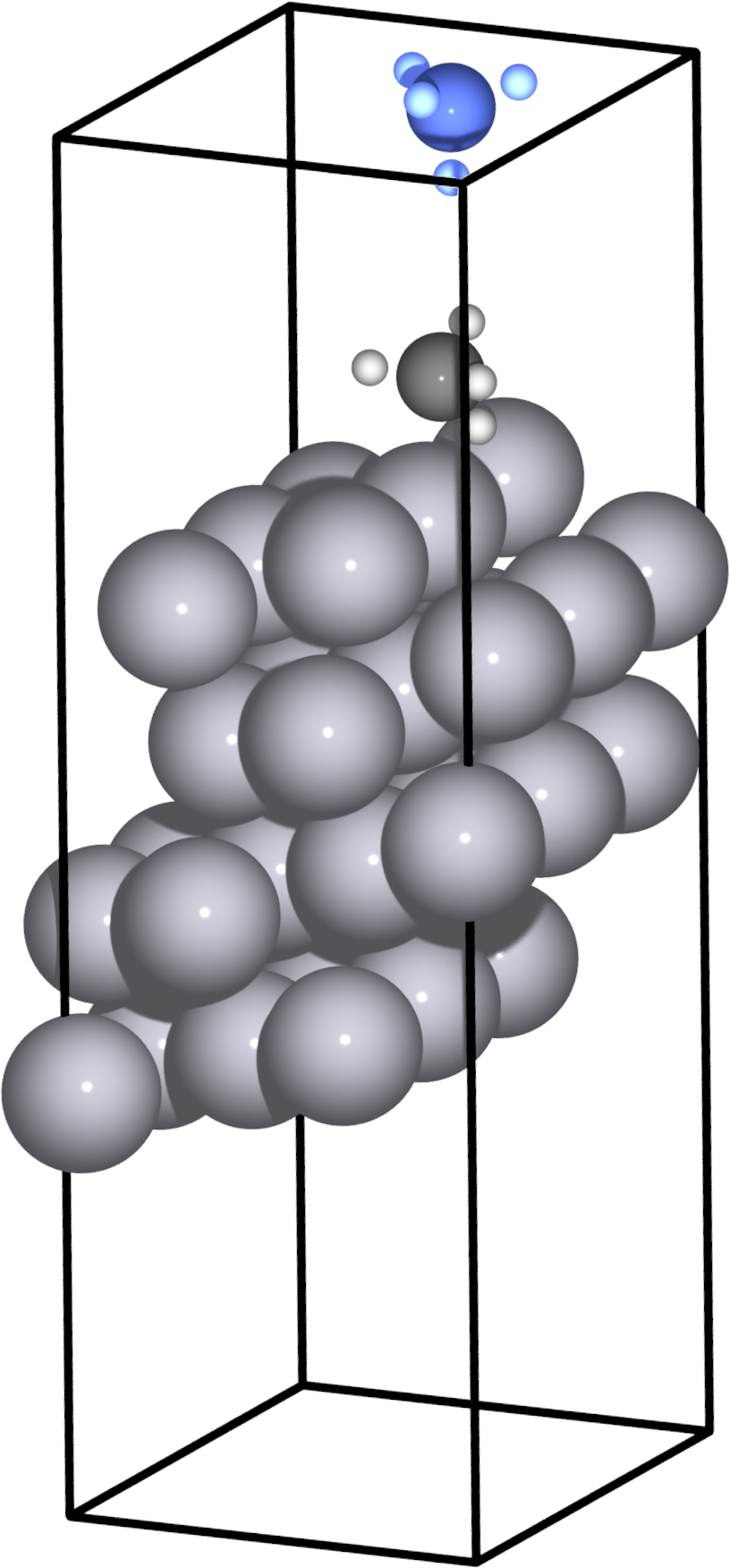}
 \caption{Depiction of slab geometries used for surface barrier height computation. Shown are both the transition state geometry (gray color) and the asymptotic state with the gas phase molecule (example considered here: CH$_4$; light blue color) far from the metal surface (here: Pt(211)).}
\label{fig:slab}
\end{figure}

We follow a computationally efficient protocol for simulating barriers for dissociative chemisorption developed in Ref.~\citenum{SBH17}. First, the lattice constant of the transition-metal is optimized with the underlying XC functional. With this lattice spacing, a slab model representing the surface is constructed, and the surface layer spacings are relaxed. Furthermore, the gas phase geometry of the dissociated molecule is relaxed. According to the `medium' algorithm from Ref.~\citenum{SBH17}, the transition state geometry is here not relaxed, but kept at a well optimized state with a functional chosen and tested reaction-by-reaction in Ref.~\citenum{SBH17}. The barrier is then calculated as the total energy difference with the molecule placed far from the surface and the transition state configuration (see Fig.~\ref{fig:slab}). Moving the molecule far from the surface rather than expressing the desorbed state as the total energy sum of slab and isolated molecule is found to have beneficial error compensation with respect to the transition state.\cite{SBH17}

We perform the DFT simulations with the VASP code,\cite{VASP} representing ionic cores with projector-augmented wave pseudopotentials.\cite{VASPPAW} Plane-wave basis set cut-offs of 500 eV are employed. Brillouin zone sampling is performed with $k$-point spacings of $\lesssim$0.02~\AA$^{-1}$ and with Gaussian occupation broadening with a width of 50~meV. As XC functionals, we employ the surface chemistry-optimized and constrained meta-GGAs MCML\cite{MCML} and VCML-rVV10.\cite{VCML}

\section{Results}

\begin{table}
\caption{Surface barrier heights (in eV) computed using the MCML and VCML-rVV10 functionals and corresponding SBH17 reference values. The last row shows the mean-absolute error (MAE) with respect to the SBH17 dataset.}
\centering
\begin{tabular}{l d{3} d{3} d{3}}
\hline
System & \multicolumn{1}{l}{\textrm{MCML}} & \multicolumn{1}{l}{\textrm{VCML-rVV10}} & \multicolumn{1}{l}{\textrm{SBH17 reference}} \\
\hline
CH$_4$/Ir(111) & 0.791 & 0.712 & 0.836 \\
CH$_4$/Ni(100) & 0.641 & 0.471 & 0.760 \\
CH$_4$/Ni(111) & 0.995 & 0.961 & 1.015 \\
CH$_4$/Ni(211) & 0.777 & 0.977 & 0.699 \\
CH$_4$/Pt(111) & 0.759 & 0.721 & 0.815 \\
CH$_4$/Pt(211) & 0.428 & 0.386 & 0.559 \\
CH$_4$/Ru(0001) & 0.891 & 0.837 & 0.800 \\
H$_2$/Ag(111) & 1.153 & 1.227 & 1.082 \\
H$_2$/Cu(100) & 0.530 & 0.612 & 0.740 \\
H$_2$/Cu(110) & 0.900 & 1.009 & 0.789 \\
H$_2$/Cu(111) & 0.392 & 0.475 & 0.628 \\
H$_2$/Ni(111) & 0.006 & -0.002 & 0.024 \\
H$_2$/Pt(111) & 0.018 & 0.025 & -0.008 \\
H$_2$/Pt(211) & -0.039 & -0.054 & -0.083 \\
H$_2$/Ru(0001) & 0.017 & 0.008 & 0.004 \\
N$_2$/Ru(0001) & 1.668 & 1.600 & 1.840 \\
N$_2$/Ru(10$\bar 1$0) & 0.125 & 0.025 & 0.400 \\
\hline
MAE & 0.101 & 0.141 & \\
\hline
\end{tabular}
\label{tab:barriers}
\end{table}

The computed barriers for the 17 dissociative chemisorption systems in the SBH17 dataset\cite{SBH17} are presented in Table~\ref{tab:barriers} Both MCML and VCML-rVV10 overall yield relatively good results for surface barrier heights with mean-absolute errors of about 0.1~eV and about 0.14~eV, respectively, which is not chemically accurate, but as accurate as DFT ideally predicts chemical reaction energies in terms of per atom errors when the underlying XC functional performs well for the system in question. Both functionals show the largest error with an underestimation of the dissociation barrier of N$_2$ on Ru(10$\bar 1$0), by $\sim$0.3~eV and $\sim$0.4~eV for MCML and VCML-rVV10, respectively. This reaction is generally difficult to simulate with some functionals even erroneously predicting barrierless dissociation.\cite{SBH17}

\begin{figure}
 \centering
        \includegraphics[width=0.8\textwidth]{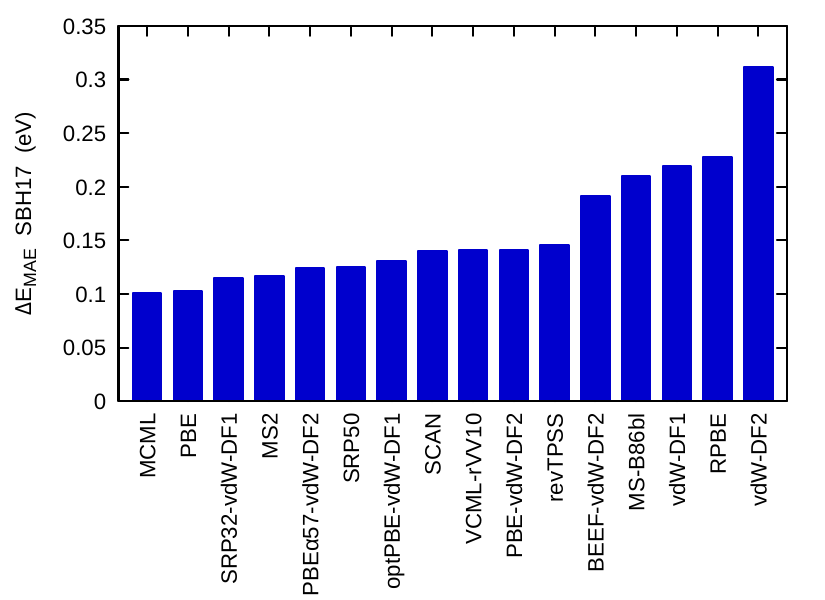}
 \caption{Different XC functionals with and without non-local correlation of vdW-DF1,\cite{VDWDF1} vdW-DF2,\cite{VDWDF2} and rVV10\cite{RVV10,VV10} type (MCML,\cite{MCML} PBE,\cite{PBE} SRP32-vdW-DF1,\cite{SBH17} MS2,\cite{MS2} PBE$\alpha$57-vdW-DF2,\cite{PBEalpha} SRP50,\cite{SRP} optPBE-vdW-DF1,\cite{OPTPBE} SCAN,\cite{SCAN} VCML-rVV10,\cite{VCML} PBE-vdW-DF2, revTPSS,\cite{REVTPSS}, BEEF-vdW-DF2,\cite{BEEFVDW} MS-B86bl,\cite{MSB86bl} vdW-DF1,\cite{VDWDF1} RPBE,\cite{RPBE}, and vdW-DF2\cite{VDWDF2}) ranked by mean-absolute errors (MAE) with respect to the SBH17\cite{SBH17} dataset. MCML and VCML-rVV10 results: this work; all other MAEs from Ref.~\citenum{SBH17}.}
\label{fig:surfacebarriers}
\end{figure}

Comparing the mean-absolute errors across the 17 systems to the other XC functionals tested in Ref.~\citenum{SBH17}, one finds that MCML performs particularly well (as well on average as PBE; see Fig.~\ref{fig:surfacebarriers}). Surprisingly, XC functionals that are well-known to perform particularly well for chemisorption on transition-metal surfaces (RPBE\cite{RPBE}) and chemi- and physorption (BEEF-vdW-DF2\cite{BEEFVDW}) turn out to display the largest errors for surface barriers. BEEF-vdW-DF2 is entirely data-based and fulfills no exchange constraints related to the homogeneous electron gas.\cite{BEEFVDW} Its regularization penalizes higher reduced-density gradient powers in the exchange enhancement, but the local density approximation limit, {\it e.g.}, as a constant is overshot by $\sim$3\%. The MCML and VCML-rVV10 functionals with similar training data sets as BEEF-vdW-DF2 do not lead to as low chemisorption energy errors for late transition-metals,\cite{VCML} but the incorporation of physics constraints appears here to avoid an overfit to surface binding energies and lead to functionals with competitive accuracy for surface barriers despite not being explicitly optimized for them. It is remarkable, that PBE as a general-purpose GGA is among the best XC functionals on average for predicting barriers for dissociative chemisorption.

\section{Summary}

Barriers for dissociative chemisorption on transition-metals were computed with empirical XC functionals fulfilling constraints related to the free electron gas important for predictions of bulk solids. It is found that these constraints yield DFT-optimal predictions of surface barrier heights, while purely empirical XC functionals optimized for surface binding energies turn out to yield significantly larger errors due to overfitting to thermodynamic ground states. Our work demonstrates that satisfying exact physical conditions in form of the local density approximation limit and exchange gradient expansion acts as an essential regularizer for empirical XC functional training.

\section*{Acknowledgment}
This research was supported by the U.S.\ Department of Energy, Office of Science, Office of Basic Energy Sciences, Chemical Sciences, Geosciences, and Biosciences Division, Catalysis Science Program to the Ultrafast Catalysis FWP 100435, at SLAC National Accelerator Laboratory under Contract Grant No.\ DE-AC02-76SF00515.

\bibliographystyle{apsrev4-2}
\bibliography{main}

\end{document}